\documentclass[prl,twocolumn,amsmath,amssymb]{revtex4}
\usepackage{graphicx}
\usepackage{bm}
\usepackage{amsmath,amssymb}
\usepackage{color}
\usepackage{ascmac}

\begin{document}
\title{Minimum heat dissipation in measurement-based quantum computation}  
\author{Tomoyuki Morimae}
\email{morimae@gunma-u.ac.jp}
\affiliation{ASRLD Unit, Gunma University,
1-5-1 Tenjin-cho Kiryu-shi Gunma-ken, 376-0052, Japan}

\date{\today}
            
\begin{abstract}
We show that at least $2kT\ln2$ of heat dissipation per qubit 
occurs in measurement-based quantum computation
according to Landauer's principle.
This result is derived by using only the fundamental fact that
quantum physics respects the no-signaling principle.
\end{abstract}

\maketitle  
The search for faster, smaller, and more economical
computers is the central research subject in today's ever growing
digital society.
Heat generation (or equivalently energy consumption) 
during computation is one of the huge obstacles to above goals,
and in fact it has been a long standing research topic
in the interdisciplinary field between information and 
thermodynamics~\cite{Landauer,Berut,Szilard,Bennett1,Bennett2,Brillouin,
Shizume,Maruyama,Bennett_demon,Sagawa}.
In 1961, Landauer showed 
in his seminal paper~\cite{Landauer}
that an irreversible process, such as an erasure of data in a memory, 
inevitably causes a 
minimum amount of heat dissipation.
More precisely, so called Landauer's principle says that an erasure
of a single bit of information in a memory causes at least $dS=k\ln2$ of entropy
generation (or, in other words, $\delta Q=kT\ln2$ of
heat dissipation or energy consumption, since $\delta Q=dST$),
where $k$ is the Boltzmann constant and $T$ is the temperature
of the environment.
Landauer's principle has been used 
not only to exorcise Maxwell's demon in thermodynamics~\cite{Bennett_demon} 
but also to establish fundamental limits of heat generation and energy
consumption in irreversible computation.

Although heat dissipation in quantum computers has not yet been fully
studied, it will be a crucial problem as well
in a near future when we can access small realistic quantum computers.
Due to the intrinsic decoherence nature
of quantum computers, the importance of studying
minimum heat dissipation in quantum computation 
will be arguably more emphasized than that in classical computation.
The purpose of this paper is to study minimum heat dissipation
in measurement-based quantum computation (MBQC)~\cite{MBQC}.
We show that at least $2kT\ln2$ of heat dissipation per qubit
occurs in MBQC according to Landauer's principle 
(or an inequality due to Sagawa and Ueda~\cite{Sagawa}). 
Interestingly, this result is independent from any specific 
physical implementation of MBQC, and in fact derived by using only the fundamental fact
that quantum physics respects the no-signaling principle~\cite{Popescu}.
We will also see that
MBQC with the cluster state~\cite{MBQC} already achieves
this minimum heat dissipation limit,
and therefore our result is an achievable lower bound.

Our result also exorcises Maxwell's demon in MBQC.
Since MBQC uses adaptive measurements, it is a kind of a feed-back
controlled system with a demon. However, no result was obtained about the study
of MBQC from the view point of the feed-back control.
The output of MBQC
is a (classical or quantum) element extracted from many possibilities,
and therefore MBQC is an entropy decreasing process.
However, the entire system (i.e., the resource state of MBQC,
the measuring apparatus, the classical computer necessary
for the feed-forwarding, and the environment) 
is a closed system,
and therefore the entropy decrease must be compensated by 
an entropy increase of another degrees of freedom;
otherwise the second law of thermodynamics is violated.
Such a Maxwell's demon problem in MBQC is solved by
our result: an entropy increase necessarily occurs
in MBQC (because of the classical memory requirement for the demon
as we will see later), and it compensates the entropy decrease caused by
the demon's computation.

\begin{figure}[htbp]
\begin{center}
\includegraphics[width=0.25\textwidth]{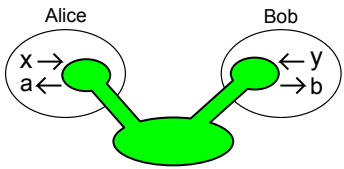}
\end{center}
\caption{
The no-signaling principle. 
} 
\label{NS}
\end{figure}

{\it No-signaling principle}.---
Let us quickly review the no-signaling principle on which
our result is based.
No-signaling principle is one of the most fundamental
principles in physics, and quantum theory also
respects it~\cite{Popescu}.
As is shown in Fig.~\ref{NS},
let us assume that Alice and Bob share a physical system, which might be
classical, quantum, or even super-quantum.
Alice chooses her measurement parameter $x$ (such as the measurement angle
of a spin), and performs measurement on her part. She obtains the result $a$.
Bob also chooses his measurement parameter $y$, and performs
measurement on his part. He obtains the result $b$.
The no-signaling principle (from Alice to Bob) is defined by
\begin{eqnarray}
P(b|x,y)=P(b|x',y)
\label{no-signaling}
\end{eqnarray}
for all $b$, $x$, $x'$, and $y$,
where $P(\alpha|\beta)$ is the conditional probability
distribution of $\alpha$ given $\beta$.
Equation~(\ref{no-signaling}) 
means that the change of Alice's measurement parameter
does not affect the probability distribution of Bob's measurement result.
In other words, the shared system cannot transmit
any message from Alice to Bob.
Interestingly, the no-signaling principle is more fundamental
than quantum theory in the sense that there is a theory
which is more non-local than quantum theory, but respects
the no-signaling principle~\cite{Popescu}.

\begin{figure}[htbp]
\begin{center}
\includegraphics[width=0.25\textwidth]{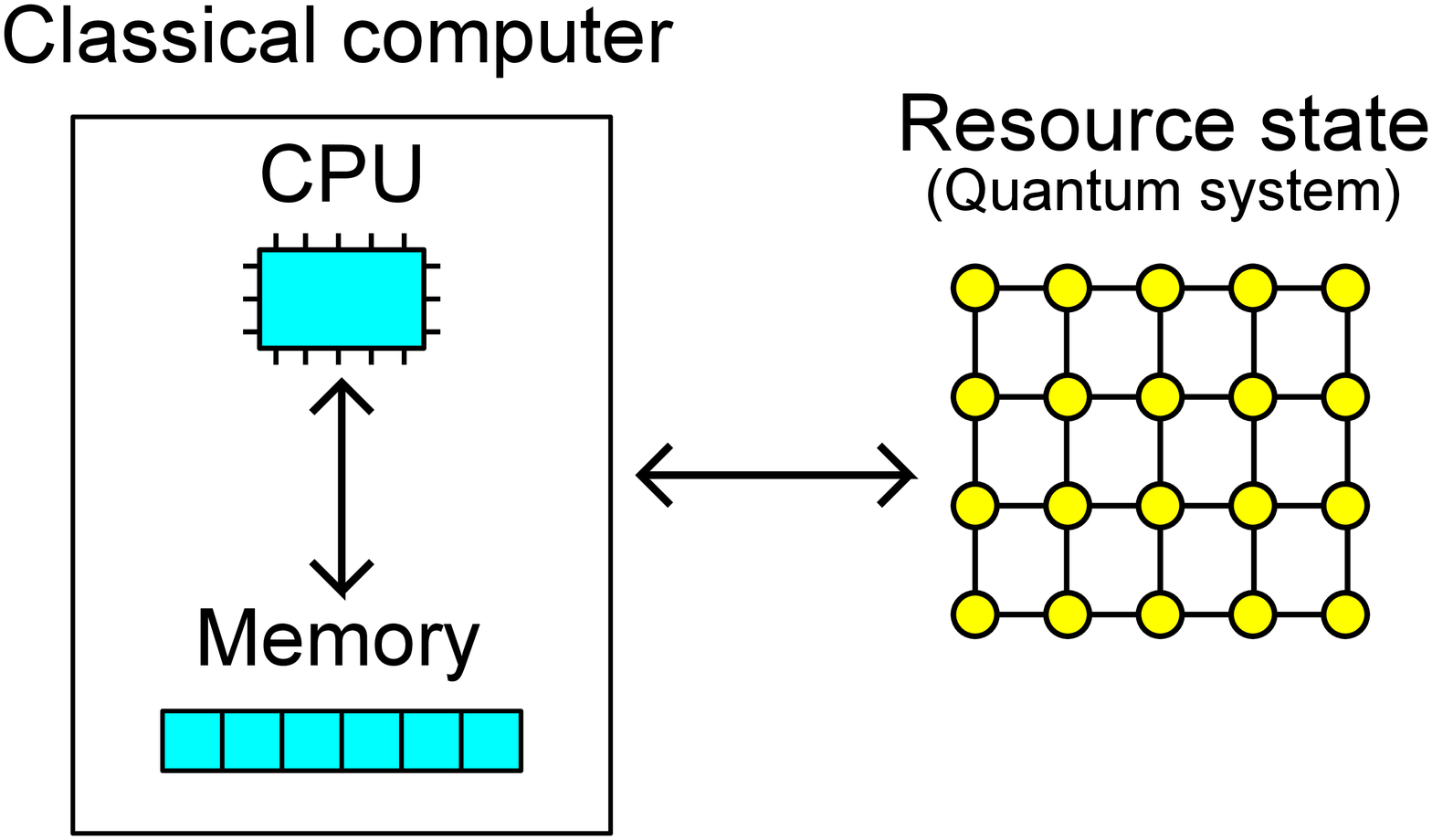}
\end{center}
\caption{
Measurement results on the resource state
are processed on a classical computer.
}
\label{CQ}
\end{figure}

{\it General MBQC}.---
Measurement-based quantum computation
is a new model of quantum computation introduced by Raussendorf
and Briegel~\cite{MBQC}.
In this model, universal quantum computation can be done
with only the preparation of a highly-entangled quantum many-body
state so called a resource state,
and adaptive local measurements on each qubit of the resource state.
Here, adaptive means that a measurement basis depends on
the previous measurement results.
Hence, in addition to the resource state, which is a quantum system,
we need a classical computer to process measurement results (Fig.~\ref{CQ}).

The computational power of MBQC is equivalent to the traditional circuit
model of quantum computation, but the clear separation between the 
quantum phase (preparation
of the resource state) and the classical phase (local adaptive
measurements) has inspired many new results which would not be
obtained if we stick to the circuit model.
For example, new resource states for MBQC which are closely connected with
condensed matter physics have been proposed~\cite{Verstraete,Gross_QCTN,
MiyakeAKLT,Cai,Miyake_edge,Miyake2dAKLT,Wei2dAKLT,Cai_magnet,fMBQC,
upload,stringnet}. 
Furthermore, a relation between MBQC and partition functions
of classical spin models were pointed out~\cite{Bravyi,Nest,Nest2}.
These discoveries have established a new bridge between
quantum information and condensed matter physics.
MBQC has also offered a new framework for fault-tolerant
quantum computing which achieves high threshold~\cite{Raussendorf_topo,
Sean,FujiiTokunaga,Ying,Ying2,FMspin}.
The quantum-classical separation in MBQC has enabled us to clarify
several relations between ``quantumness" of a resource state
and quantum computational power of MBQC on 
it~\cite{Gross_ent,Bremner,Nest_width,
Nest_preparator,Morimae_fidelity,FM_correlation}.
New protocols of secure cloud quantum computing,
so called blind quantum computing, were also developed
by using MBQC~\cite{BFK,FK,Barz,Vedran,
AKLTblind,topoblind,CVblind,topoveri,MABQC,Sueki,composable,
composableMA,distillation,Lorenzo,Joe_intern}. 

\begin{figure}[htbp]
\begin{center}
\includegraphics[width=0.25\textwidth]{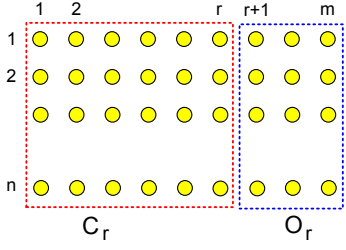}
\end{center}
\caption{
The resource state $\sigma$
for MBQC.
} 
\label{sigma}
\end{figure}

In the most general framework of MBQC, we first prepare the resource state,
$\sigma$, of $N=nm$ qubits 
as is shown in Fig.~\ref{sigma}.
Qubits are allocated
on sites of the $n\times m$ two-dimensional square lattice, and MBQC simulates
a quantum circuit with the register size $n$
and the gate depth $m-1$.
(More generally, we can consider more general graph structure
and qudits, but generalization is straightforward.
For simplicity, we consider two-dimensional square qubit system.)
Be careful that $\sigma$ is not necessarily the cluster state~\cite{MBQC}.
We do not assume any specific resource state as $\sigma$.
Measurements on all qubits in $j$th layer of $\sigma$ implement
the $n$-qubit unitary gate $U_j$.
The initial state of the computation is the $n$-qubit state $\rho_{in}$,
and it is encoded in the first layer of $\sigma$.
Let $C_r$ be the set of all qubits in the first $r$ layers of $\sigma$
(Fig.~\ref{sigma}).
We also define $O_r$, which is the set of all qubits in the last
$m-r$ layers of $\sigma$ (Fig.~\ref{sigma}).

According to the standard theory of quantum measurement~\cite{Neumann},
a measurement process on $C_r$ is described as follows.
First, an correlation between the measurement apparatus  
and the system $\sigma$ to be measured is created:
\begin{eqnarray*}
\sum_{j=1}^c
p_r^j
|j\rangle\langle j|\otimes
{\mathcal E}_r^j(\sigma),
\end{eqnarray*}
where ${\mathcal E}_r^j$ is a CPTP map,
and $p_r^j$ is a probability.
(Off-diagonal terms are omitted here for simplicity.)
Next, the projection measurement $\{|j\rangle\langle j|\}_{j=1}^c$
is performed on the apparatus,
which leads to the post-measurement state
${\mathcal E}_r^j(\sigma)$ with the probability $p_r^j$.

We require that 
\begin{eqnarray}
\rho_{out,r}^j&\equiv&
\mbox{Tr}_{C_r}[{\mathcal E}_r^j(\sigma)]
\nonumber\\
&=& 
B_r^j\Big[(U_{r}...U_{1}
\rho_{in}U_{1}^\dagger ... U_{r}^\dagger)
\otimes \eta_r^j\Big]B_r^{j\dagger},
\label{invertible}
\end{eqnarray}
where $\mbox{Tr}_{C_r}$ is the partial trace
over $C_r$, $B_r^j$ is an $(m-r)n$-qubit unitary operator,
and $\eta_r^j$ is a state of $(m-r)n-n$ qubits.
The reason why we require
the form of $\rho_{out,r}^j$ in that way
can be easily understood if we remember
that $\rho_{out,r}^j$ must contain the complete information about 
$U_r...U_1\rho_{in}U_1^\dagger...U_r^\dagger$.
(If not, we cannot proceed MBQC with $O_r$).
More precisely, it was shown in Ref.~\cite{Nayak}
that any invertible CPTP map can be written
as an application of a unitary operator on the system plus ancilla.
Therefore, the invertible CPTP map
\begin{eqnarray*}
H_n\ni U_r...U_1\rho_{in}U_1^\dagger...U_r^\dagger
\mapsto \rho_{out,r}^j\in H_{(m-r)n}
\end{eqnarray*}
has to be the form of Eq.~(\ref{invertible}),
where $H_d$ is the $d$-qubit Hilbert space.
For example, if $\sigma$ is the cluster state,
$\eta_r^j$ is the $n\times(m-r-1)$ cluster state,
and $B_r^j$ is the operation which
applies $CZ$ gates on the border between $C_r$ and $O_r$,
and random Pauli operators on $(r+1)$th layer.

In particular,
after measuring all qubits except for those in the last layer,
the state of the last layer becomes
\begin{eqnarray*}
\rho_{out,m-1}^j
= B_{m-1}^jU
\rho_{in}U
B_{m-1}^{j\dagger}
\end{eqnarray*}
with probability $p_{m-1}^j$,
where
$U\equiv U_{m-1}U_{m-2}...U_2U_1$.
The operator $B_{m-1}^j$ is an unwanted operator, so called the byproduct
operator, but we can correct it if we know $j$.
In such an MBQC, we say that our desired unitary 
$U$ is implemented
on $\rho_{in}$
up to the byproduct $B_{m-1}^j$.

\begin{figure}[htbp]
\begin{center}
\includegraphics[width=0.3\textwidth]{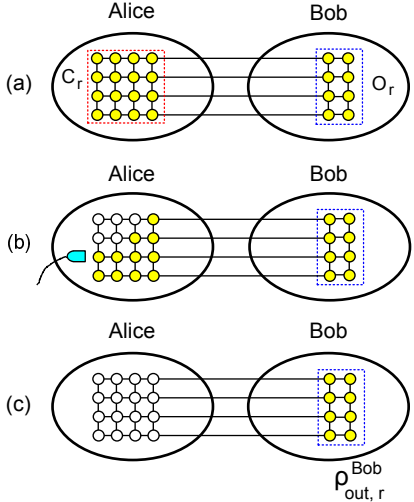}
\end{center}
\caption{
Two party general MBQC between Alice and Bob.
} 
\label{AliceandBob}
\end{figure}

{\it Minimum heat dissipation in general MBQC}.---
Now let us show that in the above general MBQC,
at least $2kT\ln2$ of heat dissipation per qubit is necessary.
In order to see it, let us consider MBQC between two party, Alice
and Bob as is shown in Fig.~\ref{AliceandBob}.
The resource state $\sigma$ is shared between Alice and Bob.
Alice possesses the subsystem $C_r$ and Bob does $O_r$.
Alice performs MBQC on her part.

The state immediately before
Alice performing the measurement on her apparatus
is
\begin{eqnarray}
\sum_{j=1}^c p_r^j |j\rangle\langle j|\otimes 
{\mathcal E}_r^j(\sigma).
\label{rhoAliceBob}
\end{eqnarray}
If we trace out Alice's system, we obtain Bob's system, 
\begin{eqnarray*}
\rho_{out,r}^{Bob}\equiv\sum_{j=1}^c p_r^j \rho_{out,r}^j.
\end{eqnarray*}

Now we point out that $\rho_{out,r}^{Bob}$ must be completely
independent from $U_r...U_1\rho_{in}U_1^\dagger...U_r^\dagger$, 
since otherwise Bob can gain some information about 
$U_r...U_1\rho_{in}U_1^\dagger...U_r^\dagger$
by measuring $\rho_{out,r}^{Bob}$.
If Bob gains some information about 
$U_r...U_1\rho_{in}U_1^\dagger...U_r^\dagger$,
it contradicts to the no-singling principle,
because
Alice can send some message to Bob by encoding her
message into $U_r...U_1\rho_{in}U_1^\dagger...U_r^\dagger$.
In Refs.~\cite{Boykin,Mosca,Nayak}, it was shown that
the entropy $H(p_r^1,...,p_r^c)$ of $\{p_r^j\}_{j=1}^c$
must satisfy
\begin{eqnarray}
H(p_r^1,...,p_r^c)\equiv-\sum_{j=1}^cp_r^j\log p_r^j\ge 2n
\label{Boy}
\end{eqnarray}
if the map 
\begin{eqnarray*}
H_n\ni \xi\mapsto
\sum_{j=1}^cp_r^jB_r^j
(\xi\otimes \eta_r^j)
B_r^{j\dagger}
\in H_{(m-r)n}
\end{eqnarray*}
works as the completely secure quantum one-time pad encryption
for any $n$-qubit state $\xi$.

Therefore, Alice has to gain at least $2n$ bit (or 
$2n\ln2$ nit in the natural base) of information
when she measures her apparatus
in Eq.~(\ref{rhoAliceBob}).
The acquisition of information is accompanied by the erase of it.
According to Landauer's principle, 
this means that at least $2nkT\ln2$ of heat is dissipated when 
the data in the memory is erased.
Therefore, we conclude that at least $2kT\ln2$ of heat dissipation
per qubit occurs in MBQC.

Landauer's principle is not
a mathematical theorem, but an ``observation" derived from
physically reasonable arguments. 
In order to obtain more precise statement,
we have to specify the model.
For example, if the erasure model is described by
the thermalization of the memory interacting with
a heat bath, Landauer's principle can be derived~\cite{Maruyama}.
Recently, Sagawa and Ueda~\cite{Sagawa} introduced
an inequality which generalizes Landauer's principle
by assuming the following memory model:
the state of the memory storing $j$th result is
given by the canonical state $\rho_{j,can}^M=\exp(-H_j^M/kT)/Z_j^M$,
where $H_j^M$ is the Hamiltonian of the memory if it stores $j$th result
and $Z_j^M\equiv\mbox{Tr}[\exp(-H_j^M/kT)]$.
The state of the memory before erasure is given by
$\sum_jp_j\rho_{j,can}^M$. 
In order to erase data in the memory,
we couple the memory with the heat bath,
$\rho_{can}^B\equiv\exp[-H^B/kT]/Z^B$,
where $H^B$ is the Hamiltonian of the bath,
and $Z^B\equiv\mbox{Tr}[\exp(-H^B/kT)]$.
The memory plus bath unitary time evolve to the final state $\rho^{MB}$.
Under these assumptions, they derived
\begin{eqnarray*}
W_{eras}+\Delta F^M\ge kTH,
\end{eqnarray*}
where 
\begin{eqnarray*}
W_{eras}&\equiv& \mbox{Tr}(\rho^{MB}(H_0^M+H^B))\\
&&-\sum_jp_j\mbox{Tr}(\rho_{j,can}^MH_j^M)-\mbox{Tr}(\rho_{can}^BH^B)
\end{eqnarray*}
is the work required for the erasure of data
in the memory, and 
$\Delta F^M\equiv kT\ln Z_0^{M}-
\sum_jp_jkT\ln Z_j^M$
is the change of the free energy of the memory
due to the erasure~\cite{IQC}.
According to Eq.~(\ref{Boy}),
we obtain $W_{eras}+\Delta F^M\ge 2nkT\ln 2$.
In particular, if we consider the case when
the memory's Hamiltonian does not depend on the stored data,
i.e., $H_j^M=H_0^M$ for all $j$,
we conclude that $W_{eras}\ge 2nkT\ln 2$,
which means that at least $2kT\ln2$ of work per qubit is necessary
in MBQC.

{\it Minimum heat dissipation in the cluster state MBQC}.---
As an example, let us consider the minimum heat 
dissipation limit
in the cluster state MBQC~\cite{MBQC}.
In the cluster state MBQC, it is well known that
measurement results for two previous layers must be
kept in order to correct byproduct operators~\cite{MBQC}.
Therefore, $2$ bit of classical memory per register qubit
is always required at any measurement step of the cluster state MBQC.
This means that $2kT\ln2$ of heat dissipation occurs at every step
of the cluster state MBQC.
In this way, the cluster state MBQC already achieves the 
minimum heat dissipation limit.
Hence we also conclude that our general limit, $2kT\ln2$,
is an achievable limit (i.e., not too under-estimating limit.)

For the cluster state MBQC, to derive the $2kT\ln2$
limit is somehow straightforward.
However, we want to emphasise here that 
our limit $2kT\ln2$ for general MBQC given in the previous section
does not make any assumption on
the resource state, the way of measurements, and the way of classical
processing. We have derived a general limit by using
only the no-signaling principle. In a future someone might find a very complicated
and drastically new MBQC far from the cluster state MBQC. 
(For example non-local many-body measurements might be allowed.)
However, our minimum heat dissipation limit $2kT\ln2$ still holds 
for such a new MBQC as long as
it satisfies the no-signaling principle.

{\it Implication for blind quantum computation}.---
Blind quantum computation is a new secure cloud quantum computing
protocol where a client (Alice), who does not have enough quantum
technology, can delegate her quantum computation to a server (Bob),
who has a full-fledged quantum computer, without leaking
any information~\cite{BFK,FK,Barz,Vedran,
AKLTblind,topoblind,CVblind,topoveri,MABQC,Sueki,composable,
composableMA,distillation,Lorenzo,Joe_intern}. 
In blind quantum computing, the no-signaling requirement
of the present result is replaced with the 
security requirement that server's state must be one-time padded. 
As is pointed out in Ref.~\cite{BFK}, the client requires only a
classical computer if she interacts with two servers.
In this case, we have only to minimize client's classical technological
requirement. Our result suggests that
the client cannot be completely free from any
technology; she has to possess at least $2$ bit of classical memory per
qubit in blind quantum computing.

{\it Implication for classical technology requirement in MBQC}.---
The requirement for classical technology in MBQC
is a research subject which has not been fully studied.
Although classical computation is cheap compared with quantum operations,
detailed understanding of the classical part is important,
because for example the latency of classical computation could 
contribute to the entire decoherence time of MBQC.
The requirement for the power of the classical CPU
was studied in Ref.~\cite{Janet}:
classical XOR gate is sufficient for universal MBQC with cluster
state, whereas
a classical universal gate set
is necessary if we use certain resource states (quantum computational
tensor-network states~\cite{Gross_QCTN}) in stead of the cluster 
state. 
Our result clarifies the
minimum classical memory requirement, namely two bits per qubit, for MBQC.

{\it Heat generation from other degrees of freedom}.---
As is already pointed out by Landauer himself~\cite{Landauer},
heat generation from Landauer's principle
is not a practical limit, but a fundamental one:
it is absurd to use Landauer's principle
to estimate the overall heat generation in an iPhone,
since there are many other factors which contribute to
the entire heat generation, and these contributions
are many orders of magnitude larger than that from Landauer's principle.
This is also the case for our result.
For example, measuring process could cause heat dissipation.
If a qubit is encoded in a polarization of a single photon,
or a two-level of an atom,
a photon detection is necessary for the measurement.
Usually, a photon detection generates a dissipative
electric current, which generates heat of many magnitude larger
than the Landauer limit.
However, these contributions are implementation specific,
and therefore beyond the scope of the present paper.
(For example, the specific measurement model which uses photons
by Brillouin~\cite{Brillouin} needs some energy for demon.)
The purpose of the present paper is to derive a fundamental limit
that is independent from any specific implementation,
like the original motivation of Landauer's paper~\cite{Landauer}.

TM is supported by Program to Disseminate Tenure
Tracking System by MEXT, Japan.



\begin{thebibliography}{99}
\bibitem{IQC}
Sagawa and Ueda~\cite{Sagawa} 
also derived another inequality
$W_{meas}+W_{eras}\ge kTI_{QC}$,
where $W_{meas}$ is the work
required for the measurement,
$W_{eras}$ is the work required for the erasure of the measurement result
in the memory,
and $I_{QC}$ is what they call ``QC-mutual information".
It is defined as 
$I_{QC}\equiv S(\rho)-\sum_jp_jS(\rho^j)$,
where $\rho$ is the pre-measurement state of the system,
and $\rho^j$ is the post-measurement state of the system
obtained with the probability $p_j$.
However, if $\rho$ and $\{\rho^j\}$ are all pure,
like the many cases in MBQC,
the above inequality is not useful
since $I_{QC}$ is always 0.
(For example, consider the $Z$-basis measurement on
the state $|+\rangle=|0\rangle+|1\rangle$. The measurement gets one bit
of information, whereas $I_{QC}=S(|+\rangle)-\frac{1}{2}S(|0\rangle)
-\frac{1}{2}S(|1\rangle)=0$.)
\end{thebibliography}
\end{document}